\pdfoutput=1

\documentclass[11pt,twoside,a4paper,cmspaper,final,collab]{cms-tdr}
\def\svnVersion{168102}\def\cmsCernNo{2012-253}\def\cmsCernDate{\today}\def\cmsMessage{Submitted to the Journal of High Energy Physics}

\begin{document}\cmsNoteHeader{HIG-12-010}

\hyphenation{had-ron-i-za-tion}
\hyphenation{cal-or-i-me-ter}
\hyphenation{de-vices}

\RCS$Revision: 156856 $
\RCS$HeadURL: svn+ssh://svn.cern.ch/reps/tdr2/papers/HIG-12-010/trunk/HIG-12-010.tex $
\RCS$Id: HIG-12-010.tex 156856 2012-11-07 13:55:15Z friis $
\newlength\cmsFigWidth
\setlength\cmsFigWidth{0.45\textwidth}
\ifthenelse{\boolean{cms@external}}{\providecommand{\cmsLeft}{top}}{\providecommand{\cmsLeft}{left}}
\ifthenelse{\boolean{cms@external}}{\providecommand{\cmsRight}{bottom}}{\providecommand{\cmsRight}{right}}

\cmsNoteHeader{HIG-12-010} 
\title{Search for the standard model Higgs boson produced in association with W and Z bosons in $\Pp\Pp$ collisions at $\sqrt{s}=7\TeV$}

\newcommand{\CLs}{\ensuremath{CL_\mathrm{s}}}
\newcommand{\CLb}{\ensuremath{CL_\mathrm{b}}}
\newcommand{\CLsb}{\ensuremath{CL_\mathrm{s+b}}}

\newcommand{\nanob}{\mbox{{\rm~nb}~}}
\newcommand{\fb}{\ensuremath{\mathrm{fb}}}
\newcommand{\pb}{\ensuremath{\mathrm{pb}}}
\newcommand{\ifb}{\ensuremath{\mathrm{fb^{-1}}}}
\newcommand{\ipb}{\ensuremath{\mathrm{pb^{-1}}}}
\newcommand{\grad}{\ensuremath{^{\circ}}}
\newcommand{\lsim}{\raisebox{-1.5mm}{$\:\stackrel{\textstyle{<}}{\textstyle{\sim}}\:$}}
\newcommand{\gsim}{\raisebox{-1.5mm}{$\:\stackrel{\textstyle{>}}{\textstyle{\sim}}\:$}}

\newcommand{\PZ}{\ensuremath{\mathrm{Z}}}

\newcommand{\pipm}{\ensuremath{\pi^{\pm}}}
\newcommand{\pizero}{\ensuremath{\pi^{0}}}
\newcommand{\Hi}{\PH}
\newcommand{\W}{\PW}
\newcommand{\Wjets}{\ensuremath{\PZ\mathrm{+jets}}}
\newcommand{\Zjets}{\ensuremath{\PZ/\gamma^{*}+\mathrm{jets}}}
\newcommand{\Wt}{\ensuremath{\mathrm{Wt}}}
\newcommand{\Wstar}{\ensuremath{\mathrm{W}^{*}}}
\newcommand{\Wparenthesisstar}{\ensuremath{\mathrm{W}^{(*)}}}
\newcommand{\WW}{\ensuremath{\PW^+\PW^-}}
\newcommand{\Zstar}{\ensuremath{\mathrm{Z}^{*}}}
\newcommand{\ZZ}{\ensuremath{\PZ\PZ}}
\newcommand{\WZ}{\ensuremath{\PW\PZ}}
\newcommand{\El}{\ensuremath{\mathrm{\mathrm{e}}}}
\newcommand{\Elp}{\ensuremath{\mathrm{\mathrm{e}}^{+}}}
\newcommand{\Elm}{\ensuremath{\mathrm{\mathrm{e}}^{-}}}
\newcommand{\Elpm}{\ensuremath{\mathrm{\mathrm{e}}^{\pm}}}
\newcommand{\Elmp}{\ensuremath{\mathrm{\mathrm{e}}^{\mp}}}
\newcommand{\M}{\ensuremath{\mu}}
\newcommand{\Mp}{\ensuremath{\mu^{+}}}
\newcommand{\Mm}{\ensuremath{\mu^{-}}}
\newcommand{\Mpm}{\ensuremath{\mu^{\pm}}}
\newcommand{\Mmp}{\ensuremath{\mu^{\mp}}}
\newcommand{\Tau}{\ensuremath{\tau}}
\newcommand{\Nu}{\ensuremath{\nu}}
\newcommand{\Nubar}{\ensuremath{\overline{\nu}}}
\newcommand{\Lep}{\ensuremath{\mathrm{\ell}}}
\newcommand{\Lepp}{\ensuremath{\mathrm{\ell}^{+}}}
\newcommand{\Lepm}{\ensuremath{\mathrm{\ell}^{-}}}
\newcommand{\Lprime}{\ensuremath{\Lep^{\prime}}}
\newcommand{\Prot}{\Pp}
\newcommand{\Pbar}{\Pap}
\newcommand{\PP}{\Prot\Prot}
\newcommand{\PPbar}{\Prot\Pbar}
\newcommand{\qq}{\ensuremath{\mathrm{q}\mathrm{q}}}
\newcommand{\Wtb}{\ensuremath{\W\mathrm{t}\mathrm{b}}}
\newcommand{\Top}{\ensuremath{\mathrm{t}}}
\newcommand{\Bot}{\ensuremath{\mathrm{b}}}
\newcommand{\Atop}{\cPaqt}
\newcommand{\Abot}{\cPaqb}
\newcommand{\To}{\ensuremath{\rightarrow}}

\newcommand{\mHi}{\ensuremath{m_{\mathrm{H}}}}
\newcommand{\mW}{\ensuremath{m_{\mathrm{W}}}}
\newcommand{\mZ}{\ensuremath{m_{\mathrm{Z}}}}
\newcommand{\mll}{\ensuremath{m_{\Lep\Lep}}}

\newcommand{\ptveto}{\ensuremath{\pt^\mathrm{veto}}}
\newcommand{\ptl}{\ensuremath{p_\perp^{\Lep}}}
\newcommand{\ptlmax}{\ensuremath{p_{\mathrm{T}}^{\Lep,\mathrm{max}}}}
\newcommand{\ptlmin}{\ensuremath{p_{\mathrm{T}}^{\Lep,\mathrm{min}}}}
\newcommand{\met}{\ensuremath{\Et^{\mathrm{miss}}}}
\newcommand{\delphill}{\ensuremath{\Delta\phi_{\Lep\Lep}}}
\newcommand{\deletall}{\ensuremath{\Delta\eta_{\Lep\Lep}}}
\newcommand{\delphimetl}{\ensuremath{\Delta\phi_{\met\Lep}}}
\newcommand{\Et}{\ensuremath{E_\mathrm{T}}}
\newcommand{\delR}{\ensuremath{\Delta R}}
\newcommand{\Eta}{\ensuremath{\eta}}
\newcommand{\GAMMA}{\ensuremath{\gamma}}

\newcommand{\effsig}{\ensuremath{\varepsilon_{\mathrm{bkg}}^{\mathrm{S}}}}
\newcommand{\effnorm}{\ensuremath{\varepsilon_{\mathrm{bkg}}^{\mathrm{N}}}}
\newcommand{\Nsig}{\ensuremath{N_{\mathrm{bkg}}^{\mathrm{S}}}}
\newcommand{\Nnorm}{\ensuremath{N_{\mathrm{bkg}}^{\mathrm{N}}}}

\newcommand{\dyee}{\ensuremath{\mathrm{\Z}/\GAMMA^*\mathrm{\to e^+e^-}}}
\newcommand{\dymm}{\ensuremath{\mathrm{\Z/}\GAMMA^*\to\mu^+\mu^-}}
\newcommand{\dytt}{\ensuremath{\mathrm{\Z}/\GAMMA^* \to\tau^+\tau^-}}
\newcommand{\dyll}{\ensuremath{\mathrm{\Z}/\GAMMA^*\mathrm{\to \ell^+\ell^-}}}
\newcommand{\zee}{\ensuremath{\mathrm{\Z\to e^+e^-}}}
\newcommand{\zmm}{\ensuremath{\mathrm{\Z}\to\mu^+\mu^-}}
\newcommand{\ztt}{\ensuremath{\mathrm{\Z}\to\tau^+\tau^-}}
\newcommand{\zll}{\ensuremath{\mathrm{\Z\to \ell^+\ell^-}}}
\newcommand{\WH}{\ensuremath{\W\Hi}}
\newcommand{\ZH}{\ensuremath{\Z\Hi}}
\newcommand{\ppww}{\ensuremath{pp \to W^+W^-}}
\newcommand{\wwlnln}{\ensuremath{W^+W^-\to \ell^+\nu \ell^-\bar{\nu}}}
\newcommand{\ww}{\ensuremath{WW}}
\newcommand{\www}{\ensuremath{\mathrm {WWW}}}
\newcommand{\wwpm}{\ensuremath{W^+W^-}}
\newcommand{\hww}{\Hi\to\WW}
\newcommand{\htt}{\Hi\to\TT}
\newcommand{\wz}{\ensuremath{WZ}}
\newcommand{\zz}{\ensuremath{ZZ}}
\newcommand{\wgamma}{\ensuremath{W\GAMMA}}
\newcommand{\wjets}{\ensuremath{W+}jets}
\newcommand{\tw}{\ensuremath{\mathrm{t}\W}}
\newcommand{\singletopt}{\ensuremath{t} ($t$-chan)}
\newcommand{\singletops}{\ensuremath{t} ($s$-chan)}
\newcommand{\emt}{\ensuremath{\Pe\Pgm\Pgt_h}}
\newcommand{\mmt}{\ensuremath{\Pgm\Pgm\Pgt_h}}

\def\fixme{({\bf FixMe})}
\newcommand{\ee}{\ensuremath{\Pe\Pe}}
\newcommand{\emu}{\ensuremath{\Pe\mu}}
\def\mm{\ensuremath{\mu\mu}}

\newcommand{\usedLumi}{5.0~\ifb}
\newcommand{\usedLumiWithSyst}{5.0~\pm~0.11~\ifb}

\date{\today}

\abstract{A search for the Higgs boson produced in association with a W or Z boson in proton-proton collisions at a center-of-mass energy of 7\TeV is performed with the CMS detector at the LHC using the full 2011 data sample, from an integrated luminosity of 5\fbinv.
Higgs boson decay modes to $\tau\tau$ and WW are explored by selecting events with three or four leptons in the final state.
No excess above background expectations is observed, resulting in exclusion limits on the product of Higgs associated production cross section and decay branching fraction for Higgs boson masses between 110 and 200\GeV in these channels.
Combining these results with other CMS associated production searches using the same dataset in the H$\to\gamma\gamma$ and H$\to \cPqb \cPaqb$ decay modes, the cross section for associated Higgs boson production 3.3 times the standard model expectation or larger is ruled out at the 95\% confidence level for a Higgs boson mass of 125\GeV.
}

\hypersetup{%
pdfauthor={CMS Collaboration},%
pdftitle={Search for the standard model Higgs boson produced in association with W and Z bosons in pp collisions at sqrt(s)=7 TeV},%
pdfsubject={CMS},%
pdfkeywords={CMS, Higgs, WH, ZH, VH, tau, WW, Associated Production}}

\maketitle

\section{Introduction}

Spontaneous electroweak symmetry breaking is introduced in the standard model (SM)~\cite{SM1,SM2,SM3} to give mass to the vector bosons ($\W^\pm$ and  $\Z$) that mediate weak interactions, while keeping the photon, which mediates electromagnetic interactions, massless.
This mechanism~\cite{Englert:1964et,Higgs:1964ia,Higgs:1964pj,Guralnik:1964eu,Higgs:1966ev,Kibble:1967sv} results in a single scalar in the SM, the Higgs boson.
While the mass of the Higgs boson is a free parameter in the SM, its couplings to the massive vector bosons, Yukawa couplings to fermions, decay branching fractions, and production cross sections in proton-proton collisions are defined and well understood theoretically~\cite{LHCHiggsCrossSectionWorkingGroup:2011ti}.
Gluon fusion (GF), weak vector boson fusion (VBF), associated production (AP) with weak bosons, and associated production with a $\ttbar$ pair ($\ttbar\Hi$) are the four most important Higgs boson production mechanisms at the Large Hadron Collider (LHC).
Although the cross section for AP is an order of magnitude lower than that of the GF mechanism, the presence of isolated high momentum leptons originating from $\W$ and $\Z$ decays suppresses the backgrounds dramatically, making these channels viable for searches for the Higgs boson.

Direct searches at the Large Electron-Positron Collider (LEP) have excluded a Higgs boson with a mass $m_{\PH} < 114.4\GeV$ at 95\% confidence level (CL)~\cite{Barate:2003sz}.
The ATLAS experiment has excluded the SM Higgs boson in the mass ranges 111--122 and 131--559\GeV~\cite{ATLASObservation}, and the CMS experiment in the mass ranges 110--121.5~\cite{CMSObservation} and 127--600\GeV~\cite{Chatrchyan:2012tx}.
Both experiments have reported the observation of a new boson with a mass near 125\GeV~\cite{ATLASObservation, CMSObservation}, predominantly in channels sensitive to Higgs bosons decaying to photon or Z boson pairs.
Tevatron experiments have reported an excess of events in the $\bbbar$ final state in the mass range 120--135\GeV~\cite{TevatronObservation}.

This paper reports a search for the SM Higgs boson produced in association with a $\W$ boson ($\WH$ channel) or a $\Z$ boson ($\ZH$ channel).
The search uses a data sample of proton-proton collisions at $\sqrt{s}=7$\TeV recorded by the Compact Muon Solenoid (CMS)~\cite{CMS-JINST} experiment at the LHC\@.
The data were collected in 2011 from an integrated luminosity of $5.00 \pm 0.11 \fbinv$~\cite{CMS-PAS-SMP-12-008}.
Throughout this document, the expression ``light lepton,'' or symbol $\ell$, will refer to an electron or muon, the symbol $\Pgt_h$ to a hadronically-decaying tau, and the symbol $L$ to an $\Pe$, $\mu$, or $\Pgt_h$.
The search for $\WH$ production is performed in three-lepton ($3L$) events in four final states with three electrons or muons ($3\ell$): $\Pe\Pe\Pe$, $\Pe\Pe\mu$, $\Pe\mu\mu$, and $\mu\mu\mu$, and two final states that have a hadronic decay of a tau ($2\ell\Pgt_h$): $\emt$ and $\mmt$.
The search for $\ZH$ production is performed in four-lepton ($4L$) events with a pair of electrons or muons consistent with the decay of a $\Z$ boson, and a Higgs boson candidate with one of the following final states: $\Pe\Pgm$, $\Pe\Pgt_h$, $\Pgm\Pgt_h$, or $\Pgt_h\Pgt_h$.
These final states can be produced by two Higgs boson decay modes: decays to a pair of $\W$ bosons ($\hww$) that both decay to leptons, and decays to a pair of taus ($\htt$).
The contribution of the $\Hi\to\PZ\PZ$ decay mode is negligible.

While the sensitivity to a Higgs boson of the AP search presented here is lower than previously published results dominated by the GF and VBF production mechanisms, the final states used in this search are essential for determining if the recently observed boson at 125\GeV is consistent with the Higgs boson predicted by the SM\@.
The Tevatron excess has been observed in the associated production $\Hi \to \bbbar$ channel~\cite{TevatronObservation}.
No evidence for associated Higgs boson production has been observed at the CMS and ATLAS experiments~\cite{ATLASObservation, CMSObservation, ATLASVHbb}.
Furthermore, the exclusive measurement of all three production processes (GF, AP, and VBF) using the $\htt$ decay mode will be critical to determine the structure of the Higgs boson couplings~\cite{Azatov:2012rd}, as the $\htt$ decay mode is the only fermionic decay mode that is experimentally sensitive to both Yukawa coupling (GF) and gauge coupling (AP and VBF) production processes.
The fermionic $\Hi\to\bbbar$ decay mode is not experimentally accessible in the GF production mechanism due to the overwhelming multijet background.

We additionally combine the searches described in this paper with previously published CMS AP searches in the $\Hi \to\gamma\gamma$~\cite{CMS-PAS-HIG-12-009} and $\Hi \to \bbbar$~\cite{Chatrchyan2012284} decay modes.
The $\Hi\to\bbbar$ result has been updated with an improved measurement of the integrated luminosity~\cite{CMS-PAS-SMP-12-008} recorded in 2011 at the CMS experiment, and this is the first time that the AP $\Hi\to\gamma\gamma$ result has been interpreted in the context of the SM\@.
With the exception of the $\Hi\to\bbbar$ search, none of the searches combined in this paper were used in the CMS observation~\cite{CMSObservation} of the new boson at 125\GeV.
This paper presents the first combination of all searches for associated Higgs boson production using the 7\TeV dataset at the CMS experiment.

\section{The CMS detector, event reconstruction, and simulation}

A more detailed description of the CMS detector can be found in Ref.~\cite{CMS-JINST}.
The central feature of the CMS apparatus is a superconducting solenoid of 6\unit{m} internal diameter, providing a field of 3.8\unit{T}.
Within the solenoid are the silicon pixel and strip trackers, which cover a pseudorapidity region of $|\eta| < 2.5$.
Here, the pseudorapidity is defined as $\eta=-\ln{[\tan{(\theta/2)}]}$, where $\theta$ is the polar angle of the trajectory of the particle with respect to the direction of the counterclockwise beam.
The lead-tungstate crystal electromagnetic calorimeter (ECAL) and the brass/scintillator hadron calorimeter (HCAL) surround the tracking volume and cover $|\eta| < 3$.
The ECAL consists of 75\,848 lead-tungstate crystals that provide coverage in pseudorapidity $\abs{ \eta }< 1.479 $ in a barrel region and $1.479 <\abs{ \eta } < 3.0$ in two endcap (EE) regions.
A preshower detector consisting of two planes of silicon sensors interleaved with a total of $3 X_0$ of lead is located in front of the EE\@.
In addition to the barrel and endcap detectors, CMS has forward calorimetry that extends the coverage to $|\eta| < 5$.
Muons are measured in gas-ionization detectors embedded in the steel return yoke, with a coverage of $|\eta| < 2.4$.

The identification of electrons, muons, and hadronically-decaying taus relies crucially on the association of tracks in the tracker with energy depositions in the ECAL for electrons, energy depositions in the HCAL for charged hadrons, and track segments in the muon system for muons.
Photons are identified as ECAL energy depositions without an associated track.
All particles are reconstructed using the particle flow (PF) algorithm~\cite{CMS-PAS-PFT-09-001}, which focuses on using an optimized combination of subdetector information to reconstruct each individual particle with the highest fidelity.
The energy resolution resulting from this reconstruction is between 1--3\% for the momentum range relevant for this analysis for electrons, photons, muons, and taus, depending on the exact
kinematics of the particular particle~\cite{CMS-PAS-EGM-10-004,MUOJINST,CMS-PAS-TAU-11-001}.

The particles reconstructed by the PF algorithm are used to construct composite objects like jets, hadronically-decaying taus, and missing transverse energy (\met), defined as the magnitude of the vector sum of the transverse momenta (\pt) of all PF objects.
The jets are identified using the anti-$k_\mathrm{T}$ jet algorithm~\cite{Cacciari:2008gp} with a distance parameter of $0.5$.
In the 2011 dataset, an average of ten interactions (pileup) occur in each proton bunch crossing.
To correct for the contribution to the jet energy due to pileup, the transverse energy density per unit area $(\rho)$ of the pileup is computed~\cite{Cacciari:fastjet1,Cacciari:fastjet2} for each event.
The energy due to pileup is estimated as the product of $\rho$ and the area of the jet, and is subtracted from the jet transverse energy (\Et)~\cite{Cacciari:subtraction}.
Subsequent to pileup subtraction, jet energy corrections are applied as a function of the jet $\Et$ and $\eta$~\cite{cmsJEC} to compensate for residual hadronic energy neglected by the jet clustering algorithm.
Hadronically-decaying taus are reconstructed using the ``hadron-plus-strips'' algorithm~\cite{CMS-PAS-TAU-11-001}, which reconstructs candidates with one or three charged pions and up to two neutral pions.

The Monte Carlo (MC) event generator \PYTHIA (version 6.424)~\cite{Pythia}  is used to generate the simulated Higgs boson samples used in this analysis.
The ZZ, WZ, and Z$\gamma$ diboson background samples are generated using {\MADGRAPH 5.1.3}~\cite{Alwall:2011uj}.
The generators use the \textsc{cteq6l}\cite{Lai:2010nw} set of parton distribution functions.
While the next-to-leading-order (NLO) calculations are used for background cross sections, the cross sections used for the Higgs boson signal samples are computed at next-to-NLO~\cite{LHCHiggsCrossSectionWorkingGroup:2011ti}.
For all processes, the detector response is simulated using a detailed description of the CMS detector, based on the \textsc{geant4} package~\cite{Agostinelli:2002hh}.
The simulations include pileup interactions matching the distribution of the number of such interactions observed in data.

\section{Trigger and event selection}
\label{sec:event_selection}

Candidate signal events are recorded if they pass a trigger requiring the presence of a high-$\pt$ electron pair, muon pair, or electron-muon pair.
The leading and subleading triggering lepton candidates are required to have $\pt > 17\GeV$ and $\pt > 8\GeV$, respectively.
Offline, electron and muon candidates are subjected to standardized quality criteria described in Ref.~\cite{egmpas} and Refs.~\cite{muonpas, Chatrchyan:2012ty}, respectively, to ensure high efficiency and precision.
In the $3L$ channels, the electron candidate is subjected to a multivariate selection exploiting the correlations among electron observables~\cite{CMS-EWK-WZ} to reduce the rate of quark or gluon jets misidentified as electrons.
Three (four) charged-lepton candidates with total charge $\pm1 (0)$ are required for the $3L$ and $4L$ channels, respectively.
The two triggering light leptons are required to have $\pt >20\GeV$ and $\pt >10\GeV$, respectively.
Non--triggering $\Pe$ and $\mu$ candidates are required to have $\pt > 10\GeV$.
The minimum $\pt$ of $\tau_h$ candidates is $20\GeV$.
Electron, muon, and $\Pgt_h$ candidates are required to originate from the primary vertex of the event, which is chosen as the vertex with highest $\sum \pt^2$, where the sum is made using the tracks associated with the vertex.
In the $4L$ channels, two leptons are required to be compatible with the decay of a $\Z$ boson, having the same flavor, opposite charge, and invariant mass within 20\GeV of the mass of the Z boson.

Leptons from the Higgs or vector-boson decays are typically isolated from the rest of the event activity, in contrast to background from jets, which are immersed in considerable hadronic activity.
For each lepton candidate a cone defined by $\Delta R \equiv \sqrt{(\Delta \eta)^2 + (\Delta \phi)^2}$, where $\phi$ is the azimuthal angle in radians, is constructed around the lepton direction at the event vertex.
The size of the cone is 0.4 for $\Pe$ and $\mu$ candidates, and 0.5 for $\tau_h$ candidates in the $2\ell\tau_h$ and $4L$ channels.
In the $3\ell$ channels a smaller $\Delta R = 0.3$ cone is used.
An isolation variable is constructed from the scalar sum of the transverse energy of all charged and neutral reconstructed particles contained within the cone, excluding the contribution from the lepton candidate itself.
The contributions of charged particles coming from pileup interactions longitudinally displaced from the primary event vertex are excluded from the isolation variable.
In the $2\ell\tau_h$ and $4L$ channels, the neutral contribution to the isolation variable from the pileup is estimated using the energy deposited by tracks from pileup vertices which point into the isolation cone, and is subtracted from the isolation variable.
In the $3\ell$ channels, the neutral contribution from pileup, which is typically composed of many low $\pt$ particle candidates, is mitigated by excluding neutral particle candidates with $\pt < 1\GeV$ from the isolation variable calculation.

For a Higgs boson mass of $125~\GeV$, the $\Hi\to\W\W\to\ell\ell$ branching fraction is approximately 1.8 times larger than the $\Hi\to\tau\tau\to\ell\ell$ branching fraction~\cite{LHCHiggsCrossSectionWorkingGroup:2011ti}. 
Accordingly, the expected signal yield in the $3\ell$ channel is dominated by the $\Hi \to \W\W$ decay.  
Conversely, the $\Hi\to\tau\tau\to\ell\tau_h$ decays dominate the signal yield in the $2\ell\tau_h$ and $4L$ channels, as their branching fraction is 3.6 times larger than the $\Hi\to\W\W\to\ell\tau_h$ branching fraction.

When the Higgs boson mass is above approximately $140\GeV$, the $\Hi \to \W\W$ decay dominates in all channels.
The topological event selections are optimized for the $\www$ final states in the $3\ell$ channels, and for the $\Hi \to \Pgt\Pgt$ final state in the $2\ell\tau_h$ and $4L$ channels.
In all channels, top-quark background events are suppressed by vetoing events containing jets with $\pt > 20\GeV$ that are identified as coming from b quarks~\cite{btag1, btag2}.
Events with additional isolated leptons ($\Pe$, $\mu$, or $\Pgt_h$ candidates) are vetoed.
In the $3L$ channels, this requirement removes diboson $\PZ\PZ \to 4\ell$ background events.
The lepton veto ensures that each channel is exclusive to all other channels presented in this paper and to the published CMS $\PH \to \PZ\PZ \to 4\ell$ analysis~\cite{PhysRevLett.108.111804}.

In the $3\ell$ channel, the dominant $\WZ \to \ell\nu\ell\ell$ background is reduced by rejecting events with a same-flavor opposite-charge lepton pair with an invariant mass within $25\GeV$ of the Z-boson mass ($m_{\Z}$).
Events are rejected if there is a jet with $\Et > 40\GeV$ to remove $\ttbar$ background events, which typically contain multiple high-$\pt$ jets.
In $\WH \to \www$ events, the neutrinos associated with the decays of the $\PW$ bosons escape detection, resulting in large \met.
Drell--Yan background events are expected to have low \met.
To mitigate degradation of the $\met$ resolution due to pileup, the minimum of two different observables is defined as the $\met$.
The first includes all PF particle candidates of the event in the computation of $\met$,  while the second uses only the charged PF particle candidates associated with the primary vertex.
To improve rejection of background events with $\met$ associated with poorly reconstructed leptons, the ``projected'' $\met$~\cite{CDF} is used.
This projected $\met$ is defined as the component of $\met$ transverse to the direction of the closest lepton if it is closer than $\pi/2$ in azimuthal angle, and the full $\met$ otherwise.
The use of both $\met$ definitions exploits the presence of a correlation between the two observables in signal events with genuine $\met$ and its absence otherwise.
Events in the $3\ell$ channel are required to have projected $\met$ above $40\GeV$.
To further reject $\W\Z$ background events, the constituents of at least one opposite-charge any-flavor (OCAF) lepton pair must be separated by less than 2 in $\Delta R$.
Finally, the smallest OCAF pair mass must be above $12\GeV$ and below $100\GeV$ to suppress $\W\gamma$ and $\W\Z$ events, respectively.

In the $2\ell\tau_h$ channels, the dominant backgrounds are $\Z$, $\W$, and $\ttbar$ events with an additional quark or gluon jet incorrectly identified as an $\Pe$, $\mu$, or $\Pgt_h$.
The probability for a quark or gluon jet to pass  the $\tau_h$ identification (misidentified $\Pgt_h$) is 10 to 100 times greater than the probability for a jet to pass the $\Pe$ or $\mu$ identification and isolation requirements.
To remove the large $\dyll +$ misidentified~$\Pgt_h$ and $\ttbar$ backgrounds, the light leptons $\Pe\mu$ ($\mu\mu$) are required to have the same charge in the $\emt$ (\mmt) channel.
The variable $L_T$, defined as the scalar sum of the transverse energy of the three lepton candidates in the event, is required to be larger than 80\GeV.
This requirement is effective in rejecting some of the background coming from the semi-leptonic decays of heavy quarks, which has a softer \pt spectrum.

The largest background in the $4L$ channels is the irreducible diboson $\Z\Z$ background.
The dominant reducible backgrounds in the $4L$ channels are $\Z + \text{2 jet}$ events, where both jets are misidentified as leptons, and $\W\Z$ events with one additional misidentified jet.
These backgrounds are highly suppressed by the lepton identification and isolation requirements.
There is an additional non--negligible contribution from $\ttbar \to \ell^+ \nu \ell^- \overline \nu \cPqb \cPaqb$ events which is suppressed by the lepton identification, isolation, and the requirement of a $\Z$-boson candidate present in the event.

\begin{figure}[!htbp]
\begin{center}
\includegraphics[width=\cmsFigWidth]{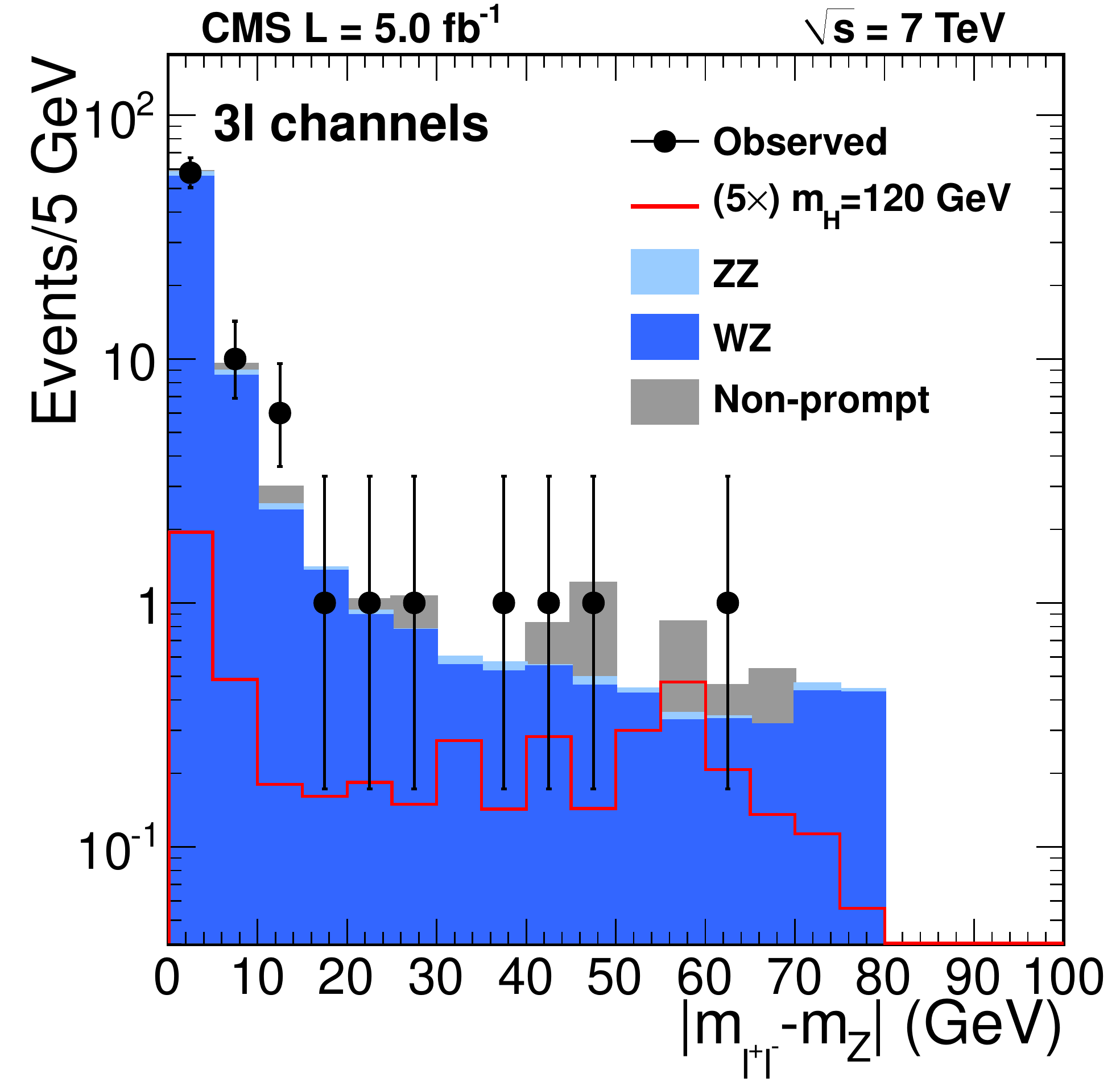}
\includegraphics[width=\cmsFigWidth]{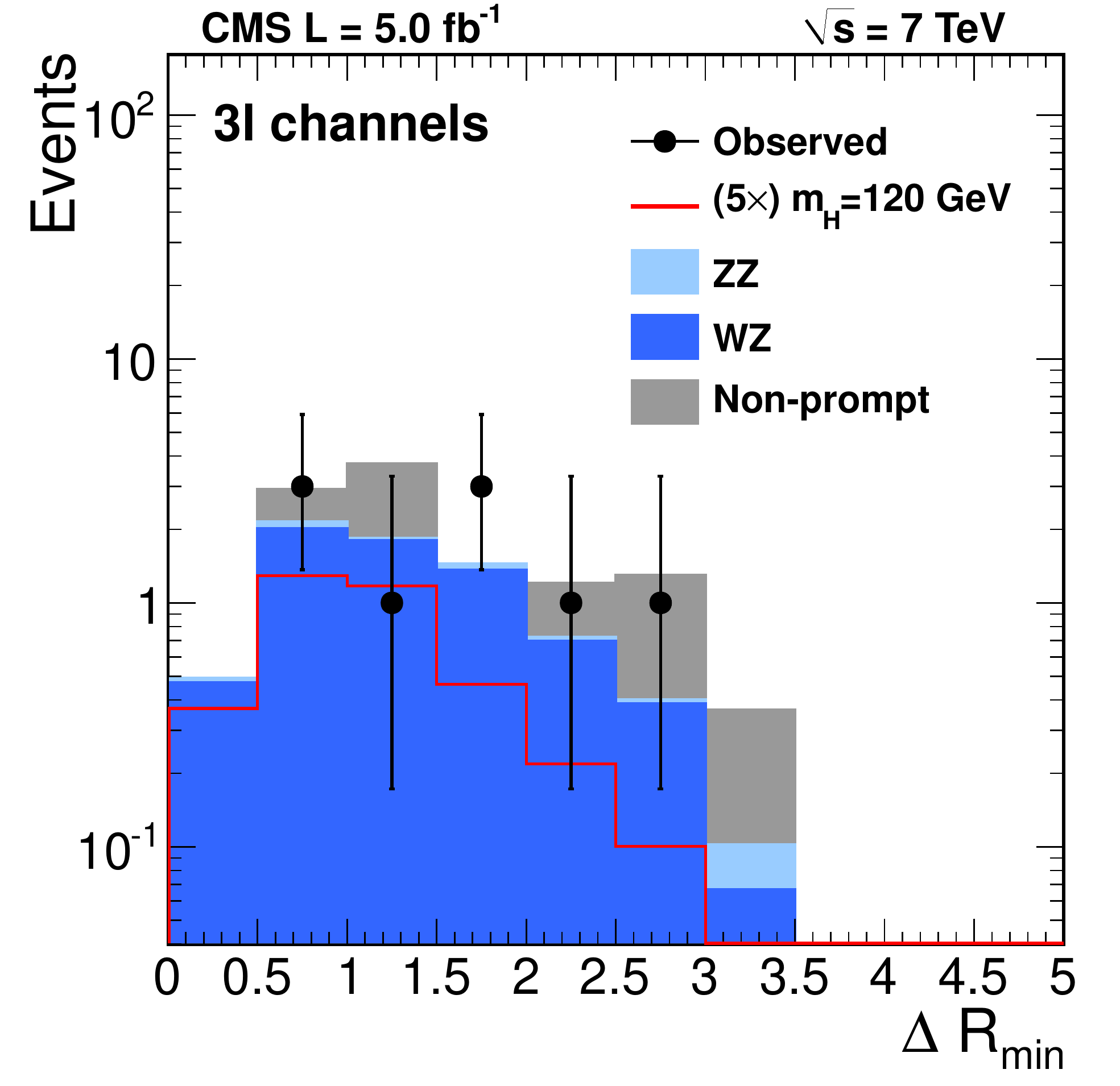}
\includegraphics[width=\cmsFigWidth]{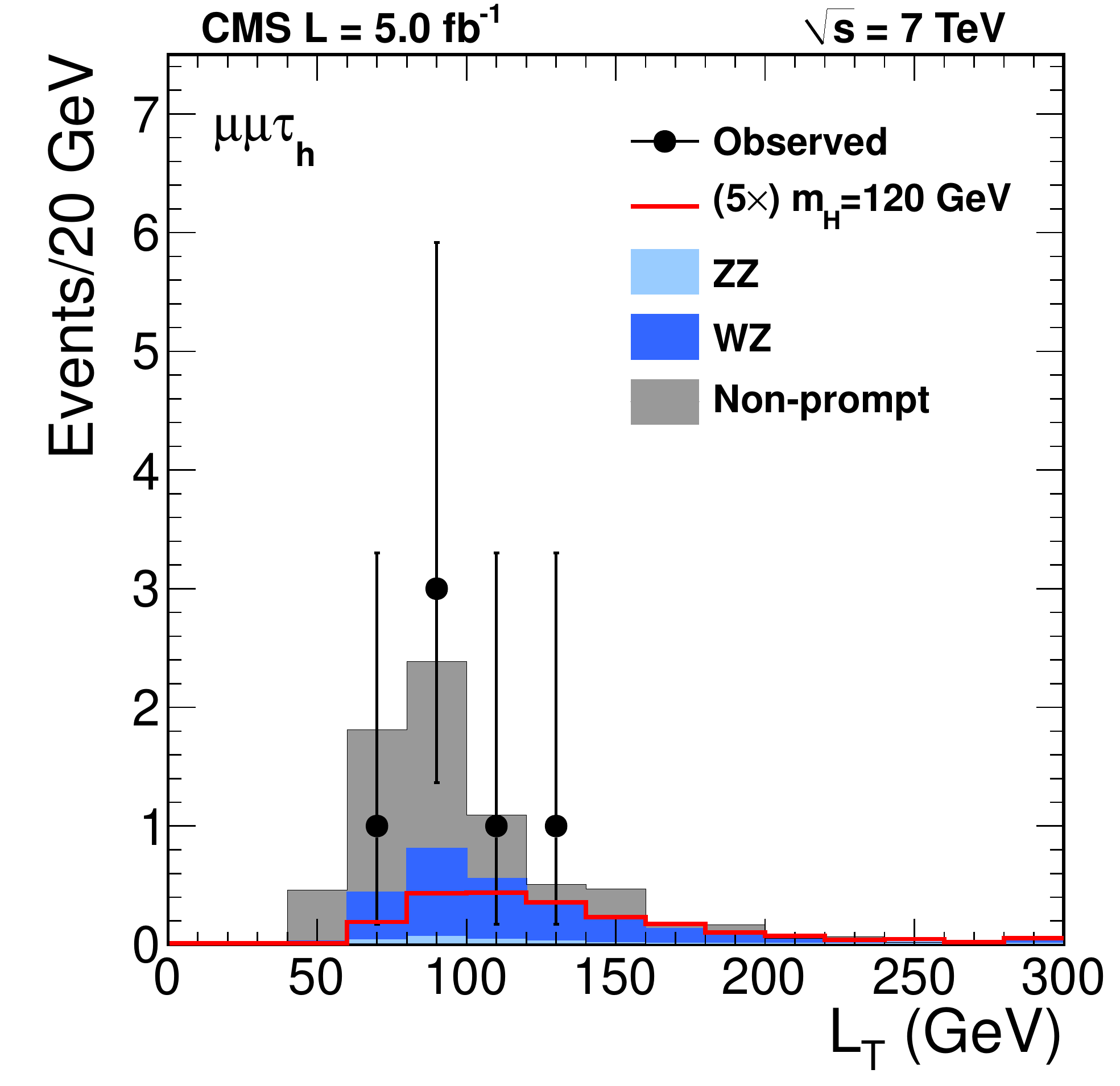}
\includegraphics[width=\cmsFigWidth]{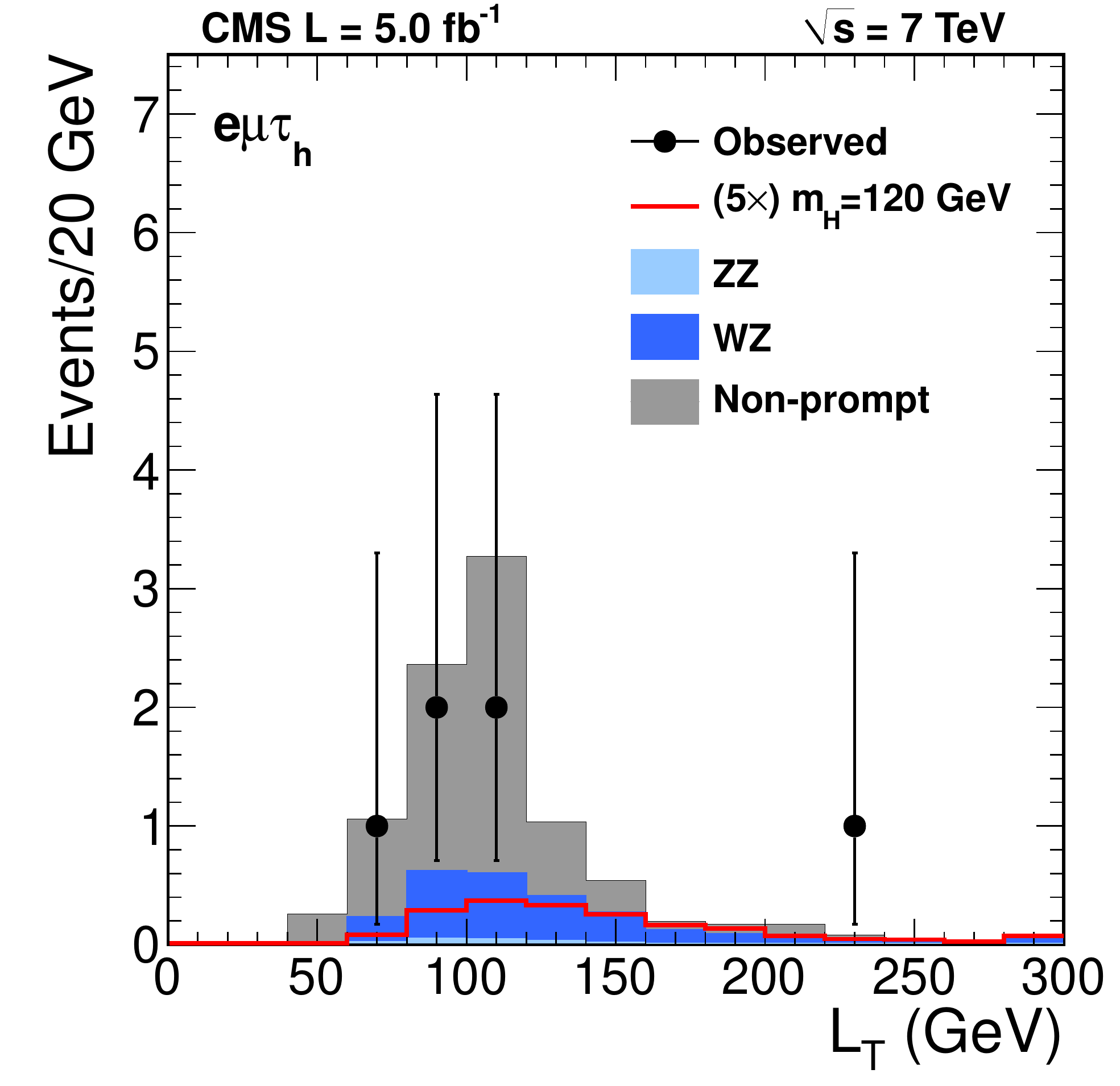}
\caption{Distributions of the dilepton mass difference with respect to $m_\Z$ in the $3\ell$ channels (upper left), the smallest $\Delta R$ distance between the opposite-charge lepton pairs in the $3\ell$ channels (upper right), $L_T$ variable in the $\mmt$ channel (bottom left), and $L_T$ variable in the $\emt$ channel (bottom right) after applying all other requirements.
The WZ, ZZ, and non-prompt backgrounds are estimated using the techniques described in Section~\ref{sec:backgrounds}.
The expected contribution from a SM Higgs boson with a mass of 120\GeV, scaled up by a factor of five, is also shown.
}
\label{fig:histo_dist}
\end{center}
\end{figure}

The resulting signal efficiencies after all selections vary between 0.1\% and 12\%, depending on production mode, decay channel, and Higgs boson mass, and are given in Table~\ref{tab:efficiencies}.
The performance of the $3\ell$ $\Z$-boson mass and minimum $\Delta R$ requirements, and the $\emt$ and $\mmt$ $L_T$ selections are illustrated in Fig.~\ref{fig:histo_dist}.
\begin{table}
  \begin{center}
    {\scriptsize
    \topcaption{Efficiency for signal events to pass the selections in each channel for the different Higgs boson production and decay modes.
    The efficiency is defined with respect to WH and ZH events in which the W or Z boson decays to final states containing an $\Pe$, a $\mu$, or a $\tau$.
    The residual corrections described in Section~\ref{sec:systematics} are applied, and the uncertainties correspond to the combined statistical and systematic uncertainties; theoretical uncertainties are not included.
    The uncertainty on the efficiency is dominated by the systematic (statistical) uncertainty for the $\Hi\to\Pgt\Pgt$ ($\Hi\to\WW$) decay in the $2\ell\tau_h$ and $4L$ channels, with the reverse being true in the $3\ell$ channels.
    \label{tab:efficiencies}
    }
    \begin{tabular} {|c|c|c|c|c|c|c|}
      \hline
      $m_\mathrm{H}$      & \multicolumn{2}{c|}{$3\ell$ channels} & \multicolumn{2}{c|}{$2\ell\tau_h$ channels} &\multicolumn{2}{c|}{$4L$ channels} \\
      & $\W\Hi\to\W\Pgt\Pgt$ & $\W\Hi\to\W\W\W$ & $\W\Hi\to\W\Pgt\Pgt$ & $\W\Hi\to\W\W\W$ & $\Z\Hi\to\Z\Pgt\Pgt$ & $\Z\Hi\to\Z\W\W$  \\
      \hline
      110 \GeV & $0.12\%\pm0.03\%$ & $2.6\%\pm0.2\%$ & $0.8\%\pm0.1\%$ & $0.2\%\pm0.1\%$ & $4.3\%\pm0.8\%$ & - \\
      120 \GeV & $0.17\%\pm0.02\%$ & $3.4\%\pm0.3\%$ & $0.9\%\pm0.1\%$ & $0.3\%\pm0.1\%$ & $4.3\%\pm0.4\%$ & $5.1\%\pm1.6\%$ \\
      130 \GeV & $0.19\%\pm0.03\%$ & $4.2\%\pm0.3\%$ & $1.1\%\pm0.1\%$ & $0.4\%\pm0.1\%$ & $4.8\%\pm0.8\%$ & $7.0\%\pm2.3\%$ \\
      140 \GeV & $0.18\%\pm0.03\%$ & $4.7\%\pm0.4\%$ & $1.2\%\pm0.1\%$ & $0.5\%\pm0.1\%$ & $5.0\%\pm0.8\%$ & $7.1\%\pm2.3\%$ \\
      150 \GeV & $0.22\%\pm0.03\%$ & $5.2\%\pm0.4\%$ & $1.4\%\pm0.1\%$ & $0.5\%\pm0.1\%$ & $4.9\%\pm0.8\%$ & $7.7\%\pm2.4\%$ \\
      160 \GeV & $0.20\%\pm0.04\%$ & $6.2\%\pm0.5\%$ & $1.6\%\pm0.1\%$ & $0.6\%\pm0.1\%$ & $5.4\%\pm0.9\%$ & $11.2\%\pm3.0\%$ \\
      \hline
    \end{tabular}
    }
  \end{center}
\end{table}

The event selections used in the $\Hi\to\gamma\gamma$ and $\Hi\to\bbbar$ channels are described in detail elsewhere~\cite{CMS-PAS-HIG-12-009, Chatrchyan2012284}.
Briefly, AP $\Hi\to\gamma\gamma$ candidate events are selected by requiring the presence of two high-$\pt$ photon candidates and an isolated electron or muon.
Events in the AP $\Hi\to\bbbar$ analysis are selected by requiring two jets identified as coming from b quarks and a vector boson candidate with high \pt.
The vector boson candidate can decay into one light lepton, two light leptons, or high $\met$, corresponding to the  $\W \to \ell \nu$, $\Z \to \ell \ell$, or $\Z \to \nu\nu$ decay modes, respectively.

\section{Background estimation}
\label{sec:backgrounds}
A combination of methods using data control samples and detailed studies with simulated events is used to estimate residual background contributions after selection.
There are two background categories: irreducible diboson backgrounds, and events with at least one non-prompt lepton.
The irreducible diboson backgrounds consist of WZ and ZZ events with the same number of isolated prompt leptons as the signal processes, and $\Z\gamma$ events with an asymmetric photon conversion.
The WZ and ZZ backgrounds are estimated using simulated samples, and are scaled by a residual correction factor obtained by comparing the observed data in diboson-enriched sidebands with the prediction from simulation.

The non-prompt lepton backgrounds arise from decays of charm and beauty quarks and ha\-drons misidentified as leptons.
The non-prompt backgrounds are evaluated using data with the ``misidentification rate method''.
The misidentification probabilities as a function of candidate $\pt$ and $\eta$, $f(\pt, \eta)$, for non-prompt lepton candidates ($\Pe$, $\mu$, or $\tau_h$) to pass the final identification and isolation criteria are measured in independent, highly pure control samples of multijet, $\W \to \mu \nu + {\rm jet}$, and $\Z \to \mu\mu + \text{jet}$ events.
The control samples are exclusive to the signal sample due to different final state topology requirements.
To minimize possible biases, the same trigger, kinematic, and quality criteria used in the final analysis are applied to the control samples.
Sidebands are defined for each channel, where all selection criteria are satisfied, with the exception that the final identification or isolation criterion is not satisfied for one or more of the final-state lepton candidates.
The sidebands are dominated by the non--prompt backgrounds.
The number of non-prompt background events in the final selection is estimated by weighting each observed non--prompt lepton candidate in the sideband by its corrected probability $f(\pt, \eta)/(1 - f(\pt, \eta))$ to pass the final identification and isolation criteria.
The estimate of the non-prompt yield in the final selection is computed using all sideband events where any two light-lepton candidates pass all requirements and the third candidate fails the isolation requirement.
In the $2\ell\tau_h$ channels, the backgrounds with a misidentified $\Pgt_h$ and two genuine prompt light leptons ($\Pe\mu$ or $\mu\mu$) are negligible, due to the requirement that the two light leptons have the same charge.
Accordingly, the misidentified-$\Pgt_h$ sideband is ignored in these channels.

Background processes with more than one non--prompt lepton, such as multijet events, $\W \to \Pgt \nu + 2\text{jet}$ in the $2\ell\tau_h$ channels, or $\Z + 2 \text{jet}$ in the $4L$ channels, are counted twice by this method since they are present in both sidebands.
The double-counting is corrected using a high-purity control region with two non-prompt leptons selected by requiring two lepton candidates to fail the isolation requirement simultaneously.
The observed events in the sideband are weighted by the corrected probability $f_1 (1-f_1)^{-1} f_2 (1 - f_2)^{-1}$, where $f_1$ and $f_2$ are the mis-identification probabilities for the leading and subleading lepton candidates, respectively, that both candidates will pass the final identification and isolation requirements; the weighted events are an independent estimate of the quantity that was double-counted.
The double-counted events are removed from the total background estimate by subtracting the independent estimate of the background with two misidentified leptons.

In the $3L$ channels, the irreducible $\WZ$ background normalization is estimated in data using a control sample of observed events with three light leptons where one of the same-flavor opposite-charge lepton pairs is compatible with a $\PZ$ boson using a ${\pm}15\GeV$ mass window.
The control sample is completely dominated by $\WZ$ events.
The same trigger and lepton identification requirements described in Section~\ref{sec:event_selection} are applied.
The $\ZZ$ background is largely reduced by the veto of events containing an additional $\Pe$, $\mu$, or $\Pgt_h$ candidate.
The theoretical NLO calculation~\cite{MCFM} is used as the normalization of the $\ZZ$ background.
The $\Z\gamma$ background, where the $\gamma$ is misidentified as an electron through an asymmetric conversion is estimated from simulation.
In the $3\ell$ channels the expected contribution from this background is negligible after the $\met$ requirement, and it is highly suppressed due to the small branching fraction in the $\tau_h$ channels.

In the $4L$ channels, $\WZ$ events have at least one non-prompt lepton and are estimated using the misidentification-rate method described above.
The dominant background comes from irreducible $\ZZ$ events.
The number of $\ZZ$ background events $N_{\PZ\PZ}^\text{est}$ is estimated by scaling the observed inclusive $\Z$ yield $N_\PZ^\text{obs}$ by the expected ratio of $\ZZ$ and $\Z$ production:
\begin{equation*}
N_{\PZ\PZ}^\text{est} = N_\PZ^\text{obs} \cdot \frac {\sigma_{\PZ\PZ}^\mathrm{SM}}{\sigma_\PZ^\mathrm{SM}} \cdot \frac {A_{\PZ\PZ}}{A_\PZ},
\end{equation*}
where $\sigma_{\PZ\PZ}^\mathrm{SM}$~\cite{MCFM} and $\sigma_{\PZ}^\mathrm{SM}$ are the theoretical SM cross sections, and $A_{\PZ\PZ}$ and $A_\PZ$ are the acceptances to pass all event selections
for the $\PZ\PZ$ and $\PZ$ processes, respectively.
The acceptances $A$ are estimated using MC simulation.
The $\Z\gamma$ background is negligible in the $4L$ channels.

\section{Efficiencies and systematic uncertainties}
\label{sec:systematics}

The trigger, identification, and isolation efficiencies for electrons and muons are measured with data using the ``tag-and-probe'' technique~\cite{CMS-EWK-WZ} in $\Z\to\ell\ell$ events.
The $\Pgt_h$ identification efficiency is measured with an uncertainty of 6\% using the tag-and-probe technique in Z$ \to \Pgt \Pgt \to \mu \Pgt_h$ events~\cite{CMS-PAS-TAU-11-001}.
Efficiencies for the Higgs boson signal and $\WZ$, $\ZZ$, and $\Z\gamma$ diboson samples are estimated using MC simulation, and residual differences between the lepton efficiencies in the simulation and data are corrected by scaling the simulation to match the efficiency measured in data.
The uncertainty on the residual correction is taken as a systematic uncertainty in the final result.
The uncertainty on the b-tagging efficiency is 6\%~\cite{Chatrchyan2012284}.
Uncertainties on the jet energy scale and $\met$ have been evaluated in $\Z$ + jet and $\gamma$ + jet events~\cite{cmsJEC}, and are propagated to systematic uncertainties on the final yields.
The uncertainty due to the pileup description is evaluated by varying the distribution of the estimated number of expected pileup interactions per event in data, and is 1\% or less.
There is a 2.2\% uncertainty~\cite{CMS-PAS-SMP-12-008} on the total integrated luminosity of the collected data sample.

Two theoretical systematic uncertainties on the overall signal yield are considered.
The uncertainty on the QCD factorization and renormalization scales affects the expected signal cross section and, in the $3\ell$ channel, the efficiency of the jet veto.
The effect of variations in the parton distribution functions, the value of $\alpha_s$, and higher-order corrections are propagated to the efficiency of the signal selection using the PDF4LHC prescription~\cite{Botje:2011sn,Alekhin:2011sk,Lai:2010vv,Martin:2009iq,Ball:2011mu}.

The methods to estimate the different backgrounds are explained in Section~\ref{sec:backgrounds}.
For the $3L$ channels, the associated uncertainty on the diboson backgrounds is 12\% and 4\% for the $\WZ$ and $\ZZ$ components, respectively.
In the $4L$ channels, the theoretical uncertainty of 10\% on the $\PZ\PZ$ production cross section~\cite{LHCHiggsCrossSectionWorkingGroup:2011ti} dominates the uncertainty on the estimate of the $\PZ\PZ$ background.
The uncertainty on the estimate of the non-prompt lepton backgrounds is 30\% and is dominated by uncertainties in the measurement of the misidentification rate.
The final estimate of the non-prompt backgrounds has an additional systematic uncertainty due to the limited number of observed events with leptons failing the isolation requirements.
In the $\emt$ and $\mmt$ mass spectra, a shape uncertainty~\cite{Conway-PhyStat} is added for each bin in the spectra, corresponding to the statistical uncertainty of the control region bin used to compute the non-prompt background estimate.

\section{Results}

After all selections, a total of 29 events are observed, while $33.5 \pm 4.3$ are expected from the background.
The number of observed and expected background events are enumerated for each channel in Table~\ref{tab:results}.
The observed data are consistent with the expected yield from the backgrounds.
The efficiency for signal events to pass all selections are detailed for each channel and Higgs boson mass, production mechanism, and decay mode in Table~\ref{tab:efficiencies}.
The efficiencies are defined with respect to events where all W and Z bosons decay to leptons (excluding $\Z \to \nu\nu$ decays).
\begin{table}[!htbp]
  \begin{center}
    {\scriptsize
    \topcaption{Observed number of events and expected number of signal and background (bkg) events for the different channels.
    The uncertainties correspond to the combined statistical and systematic uncertainty.
    The second and third columns give the expected yield of a Higgs boson signal ($m_\Hi = 120\GeV$) from the $\Hi\to\Pgt\Pgt$ and $\Hi \to \W\W $ decays, respectively.
    The theoretical uncertainties on the signal yields are not included.
    \label{tab:results}
    }
    \begin{tabular} {|c|c|c|c|c|c|c|c|}
      \hline
      Channel    & \multicolumn{2}{|c|}{SM Higgs boson (120\GeV)}  & Observed & All bkg.         & $\ZZ$            & $\WZ$            & Non--prompt bkg. \\
                 & $\Hi\to\Pgt\Pgt$      & $\Hi \to \W\W $       &      &                  & $\to 4\ell$      & $\to 3\ell$   & \\
      \hline
      \hline

      $3\ell$ & 0.13 $\pm$  0.01      & 0.55 $\pm$  0.04      & 7    & 8.45 $\pm$  1.33 & 0.27 $\pm$  0.06 & 5.65 $\pm$  0.59 & 2.52 $\pm$  1.19 \\

      \hline
      $2\ell\tau_h$       & 0.71 $\pm$ 0.06       & 0.05 $\pm$ 0.00       & 10    & 13.24 $\pm$ 2.62    & 0.38 $\pm$ 0.04   & 4.39 $\pm$ 0.60   & 8.47 $\pm$ 2.54 \\

      \hline
      $4L$      & 0.55 $\pm$ 0.06       & 0.14 $\pm$ 0.05         & 12   & 11.82 $\pm$ 2.36 & 6.04 $\pm$ 0.62    & \multicolumn{2}{c|}{5.78 $\pm$ 2.28} \\
      \hline
    \end{tabular}
    }
  \end{center}
\end{table}

\begin{figure}[!htbp]
  \begin{center}
    \includegraphics[width=\cmsFigWidth]{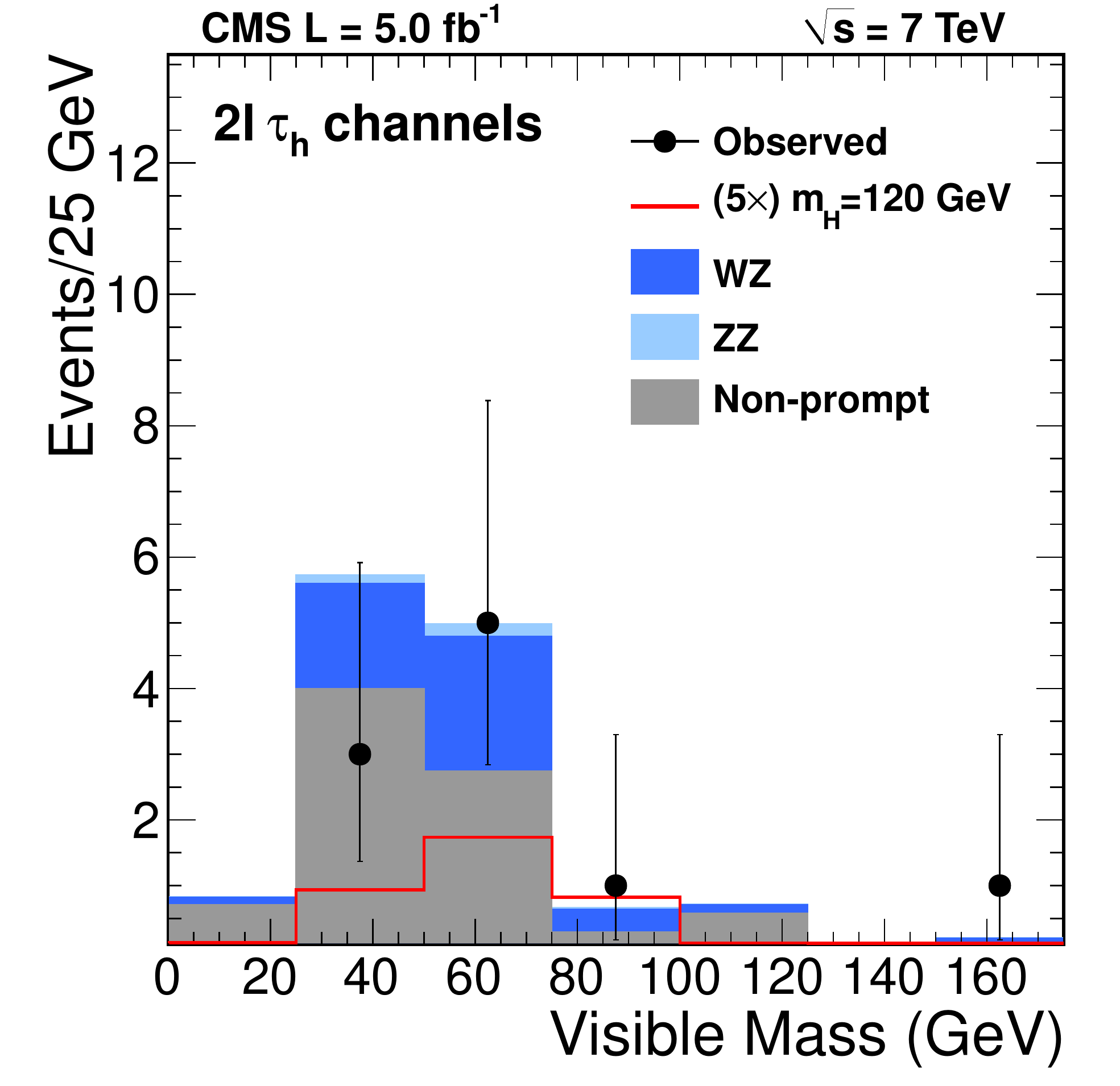}
    \includegraphics[width=\cmsFigWidth]{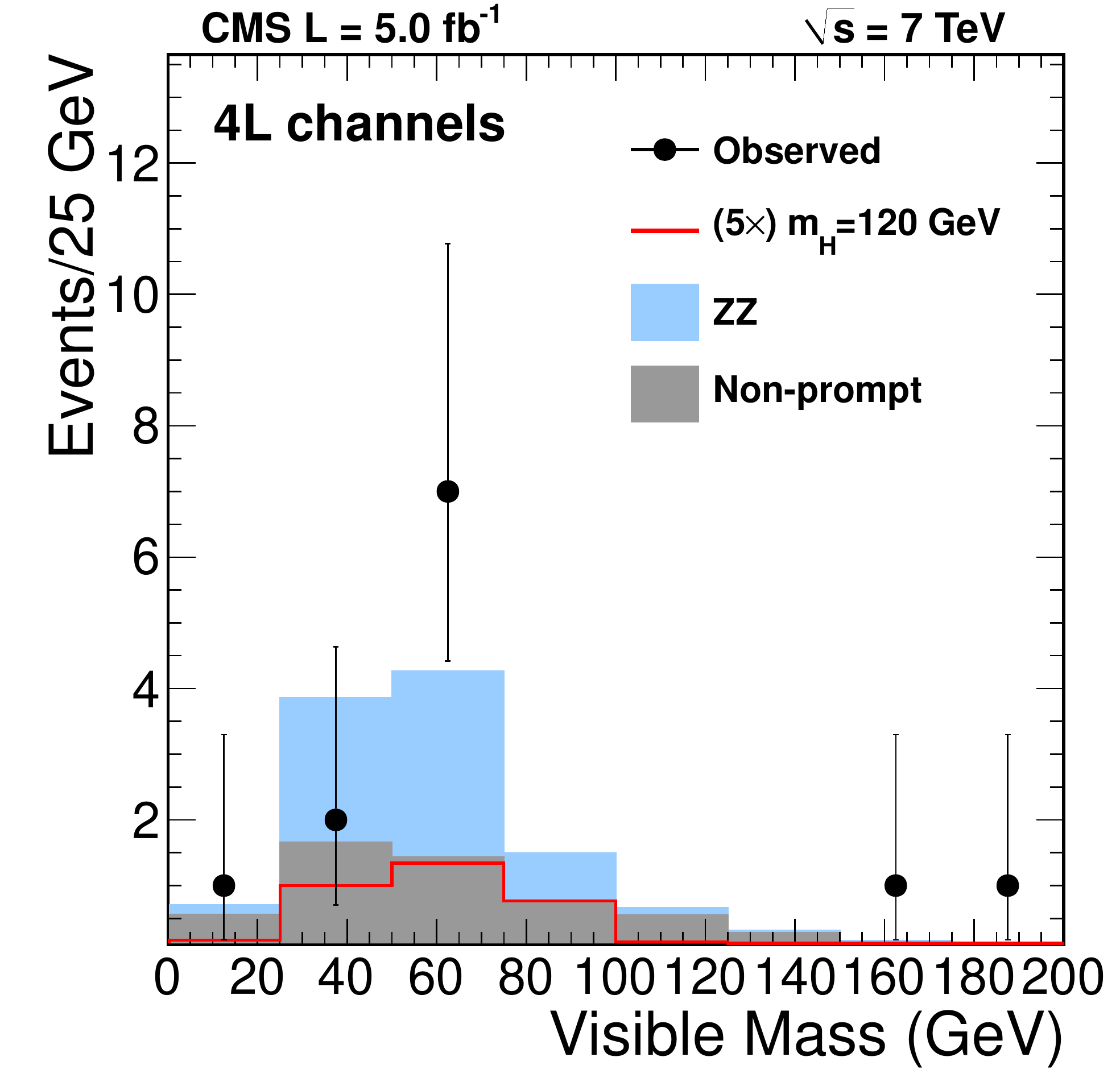}
    \caption{Visible invariant mass of the Higgs candidate in the $2\ell\tau_h$ channels (left), and $4L$ (right) channels after all selections.
    The WZ, ZZ, and non-prompt backgrounds are estimated using the techniques described in Section~\ref{sec:backgrounds}.
    The expected contribution from a SM Higgs boson with a mass of 120\GeV, scaled up by a factor of five, is also shown.
    }
    \label{fig:HiggsTauTauMassResults}
\end{center}
\end{figure}

In the $2\ell\tau_h$ channels, it is not possible to definitively assign the same--charge electrons or muons to either the $\W$ or the Higgs boson candidate.
However, as the signal is dominated by $\Hi \to \Pgt\Pgt$ decays, the final-state light leptons produced in the decays of the $\tau$ leptons have a softer \pt spectrum than light leptons from $\W \to \ell \nu$ decays, as they are associated with two neutrinos instead of one. Accordingly, we define the subleading light lepton and $\tau_h$ as the Higgs boson candidate.
The invariant mass of the Higgs boson candidate is shown for the final selected events in the $2\ell\tau_h$ and $4L$ channels in Fig.~\ref{fig:HiggsTauTauMassResults}.

\section{Limits on SM Higgs boson production}

In the searches presented in this paper, the observed events show no evidence for the presence of a Higgs boson signal, and we set 95\% CL upper bounds on the Higgs boson associated production cross section.
To obtain exclusion limits we use the \CLs\ method~\cite{Junk,Read,LHC-HCG} based on a binned likelihood of the invariant mass spectrum in the $\emt$ and $\mmt$ channels (Fig.~\ref{fig:HiggsTauTauMassResults}), and the number of observed and expected events in the $3\ell$ and $4L$ channels.
The non-prompt background mass spectra for the $2\ell\tau_h$ channels has a shape uncertainty for each bin in the spectra.
Systematic uncertainties are represented in the limit computation by nuisance parameters using a log-normal constraint.
Correlated uncertainties among channels are represented by common nuisance parameters.
The nuisance parameters are varied from one pseudoexperiment to the next in the calculation of the CL$_s$ test statistic.

\begin{figure}[!htbp]
\begin{center}
  \includegraphics[width=\cmsFigWidth]{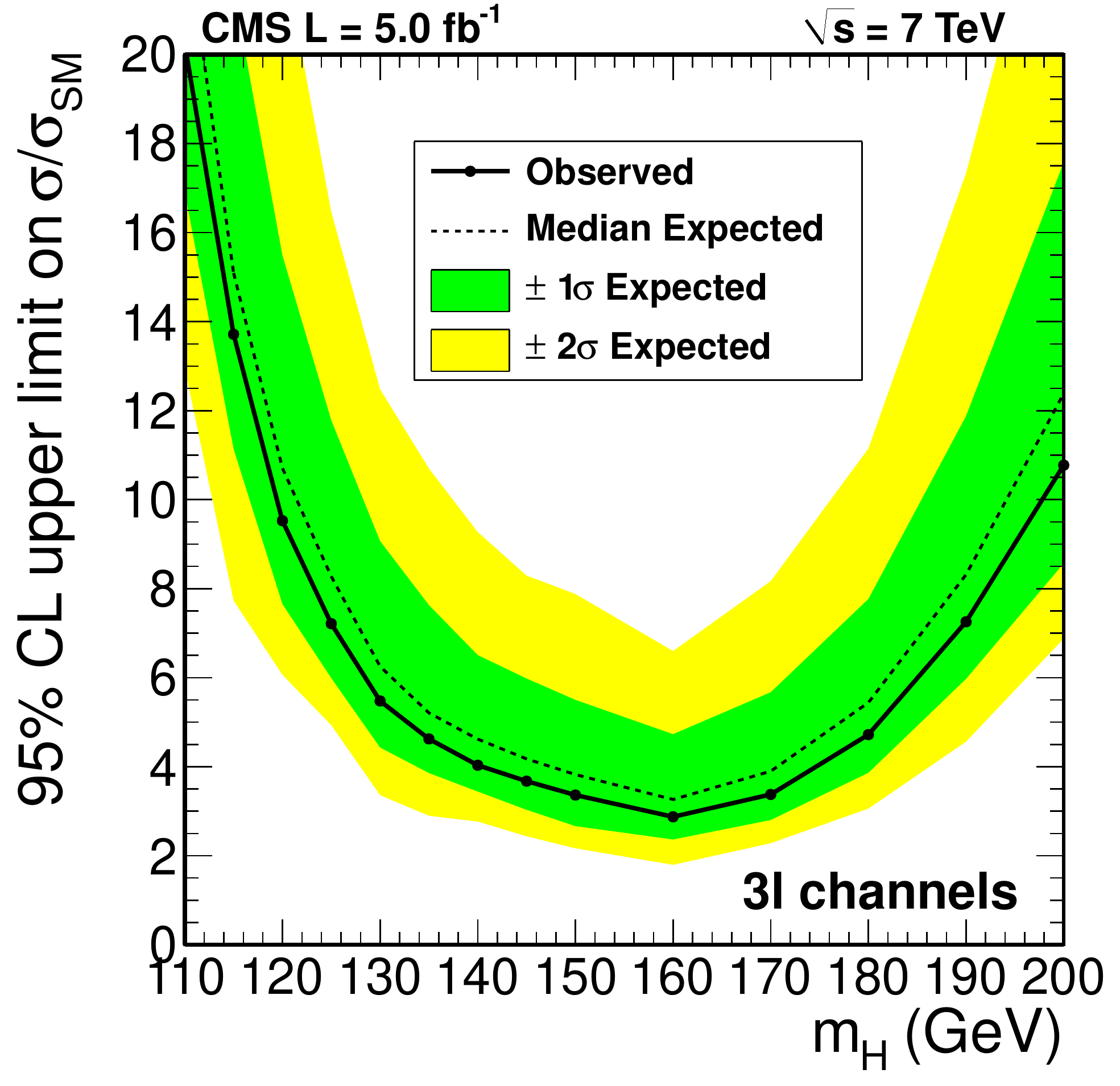}
  \includegraphics[width=\cmsFigWidth]{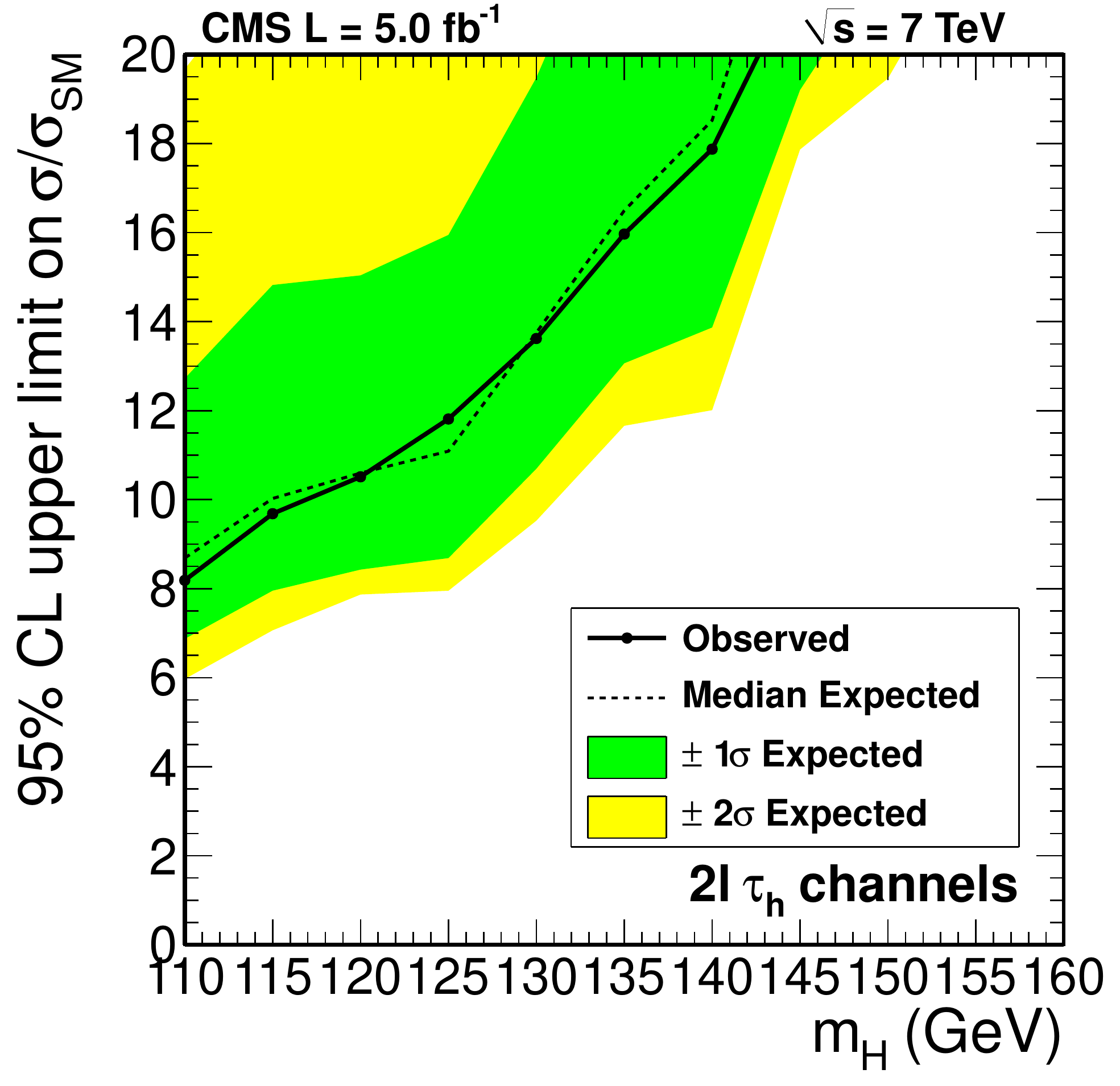}\\
  \includegraphics[width=\cmsFigWidth]{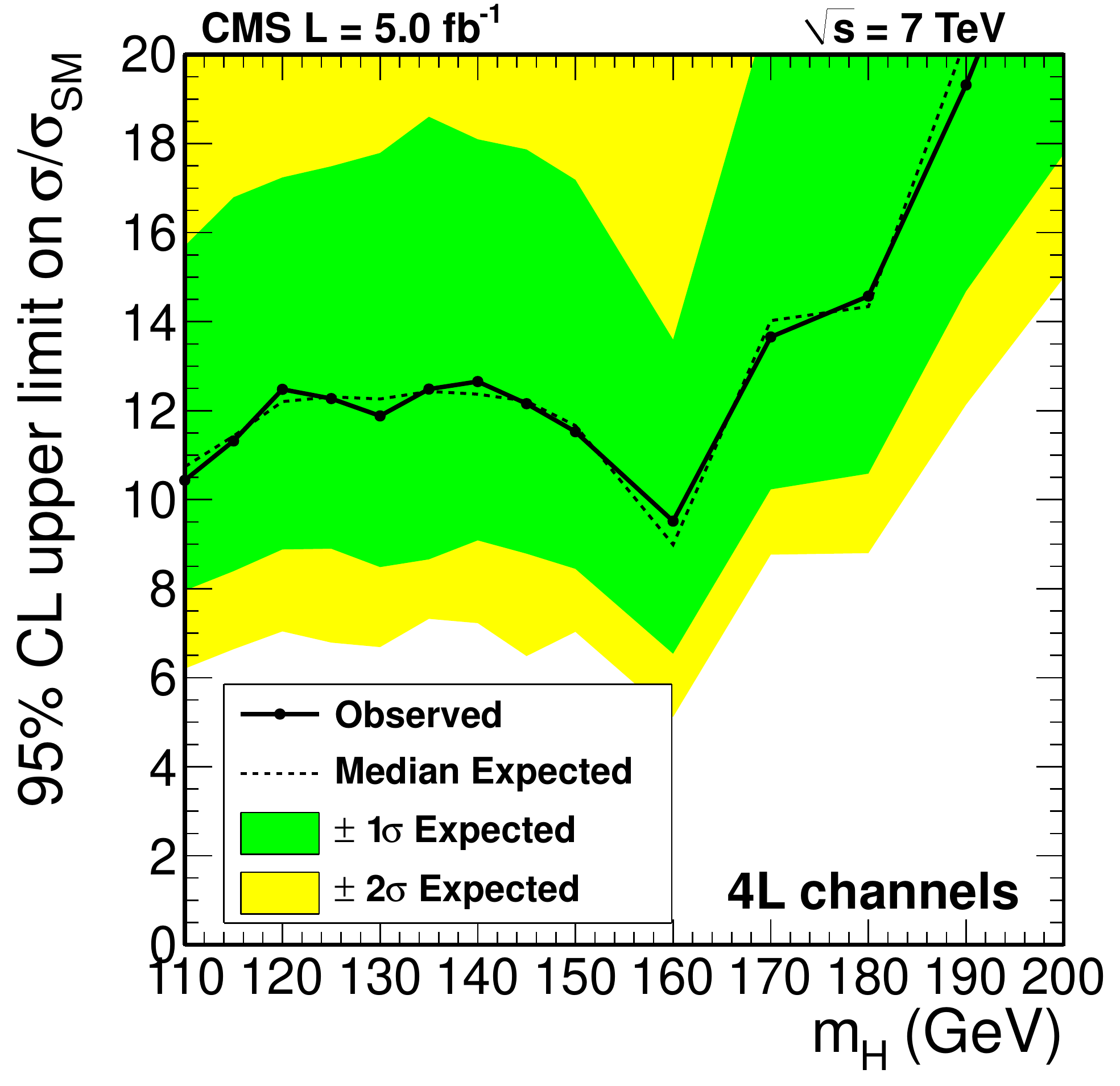}
  \includegraphics[width=\cmsFigWidth]{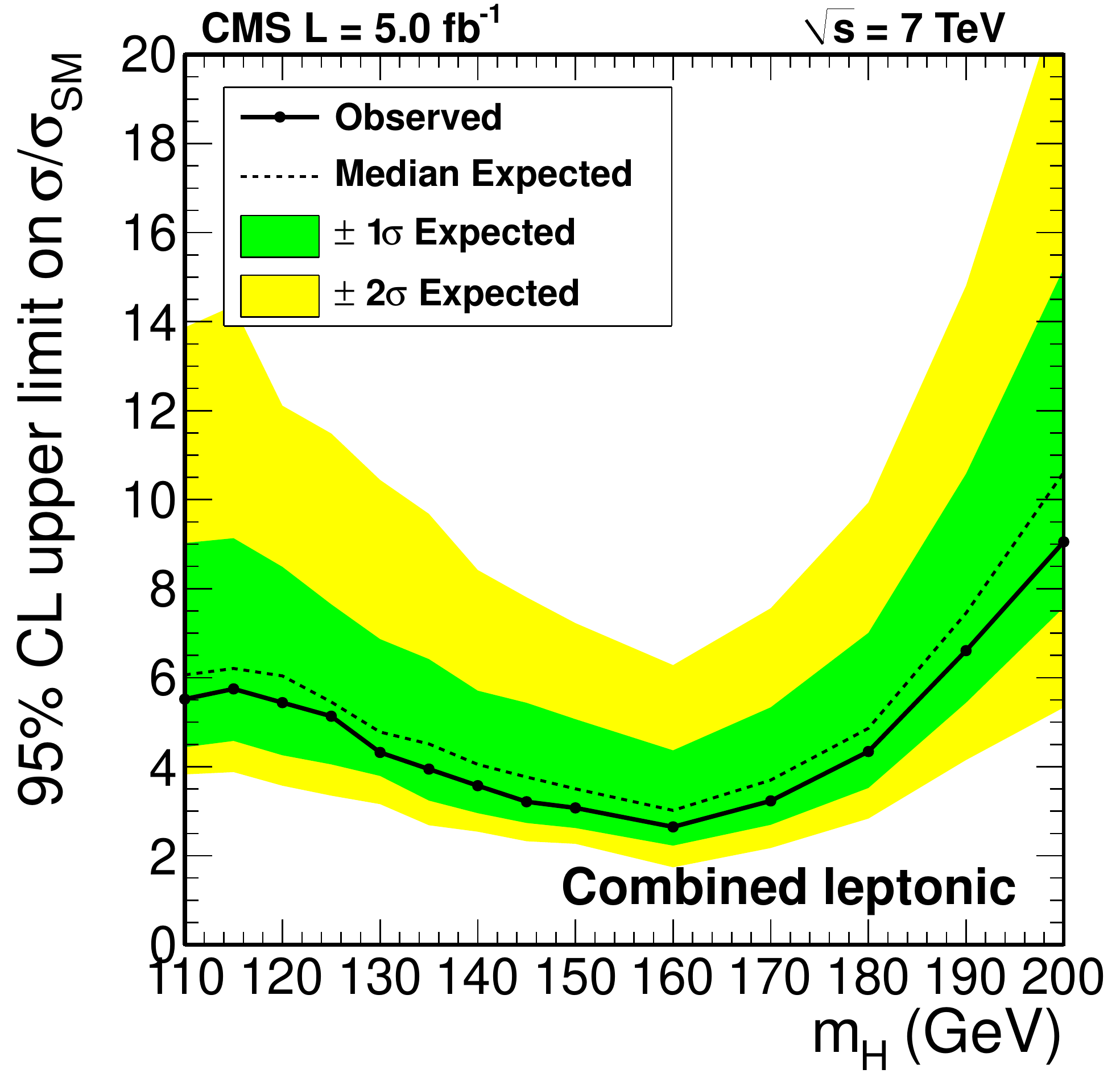}
  \caption{Observed and expected limits, at 95\% CL, on SM Higgs boson production using the $3\ell$ (top left), $2\ell\tau_h$ (top right), and $4L$ (bottom left) channels.
  The combination of the three channels is shown bottom right.}
\label{fig:combinedLimit}
\end{center}
\end{figure}

Figure~\ref{fig:combinedLimit} shows the observed and median expected 95\% CL upper limits on SM Higgs boson production set by this analysis for each channel individually and for the combination of all three.
The limit is expressed in terms of the ratio of the Higgs boson cross section times the relevant branching fractions, to that predicted in the SM, $\sigma/\sigma_\mathrm{SM}$.
The two bands give the variation around the median expected limit by one and two standard deviations.
We set a 95\% CL upper limit on $\sigma/\sigma_\mathrm{SM}$ in the range 3.1--9.1.

\begin{figure}[!htbp]
\begin{center}
  \includegraphics[width=\cmsFigWidth]{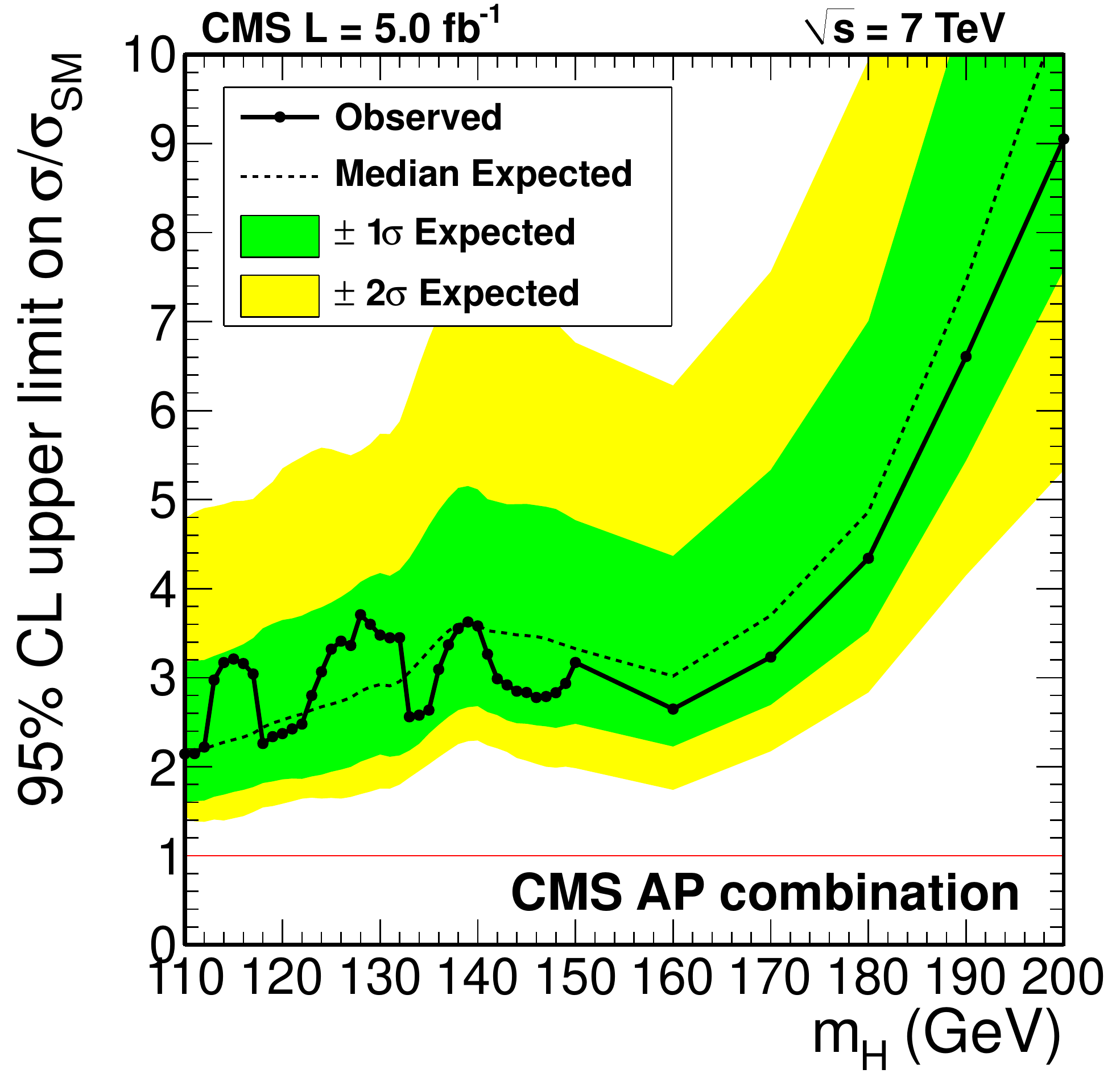}
  \includegraphics[width=\cmsFigWidth]{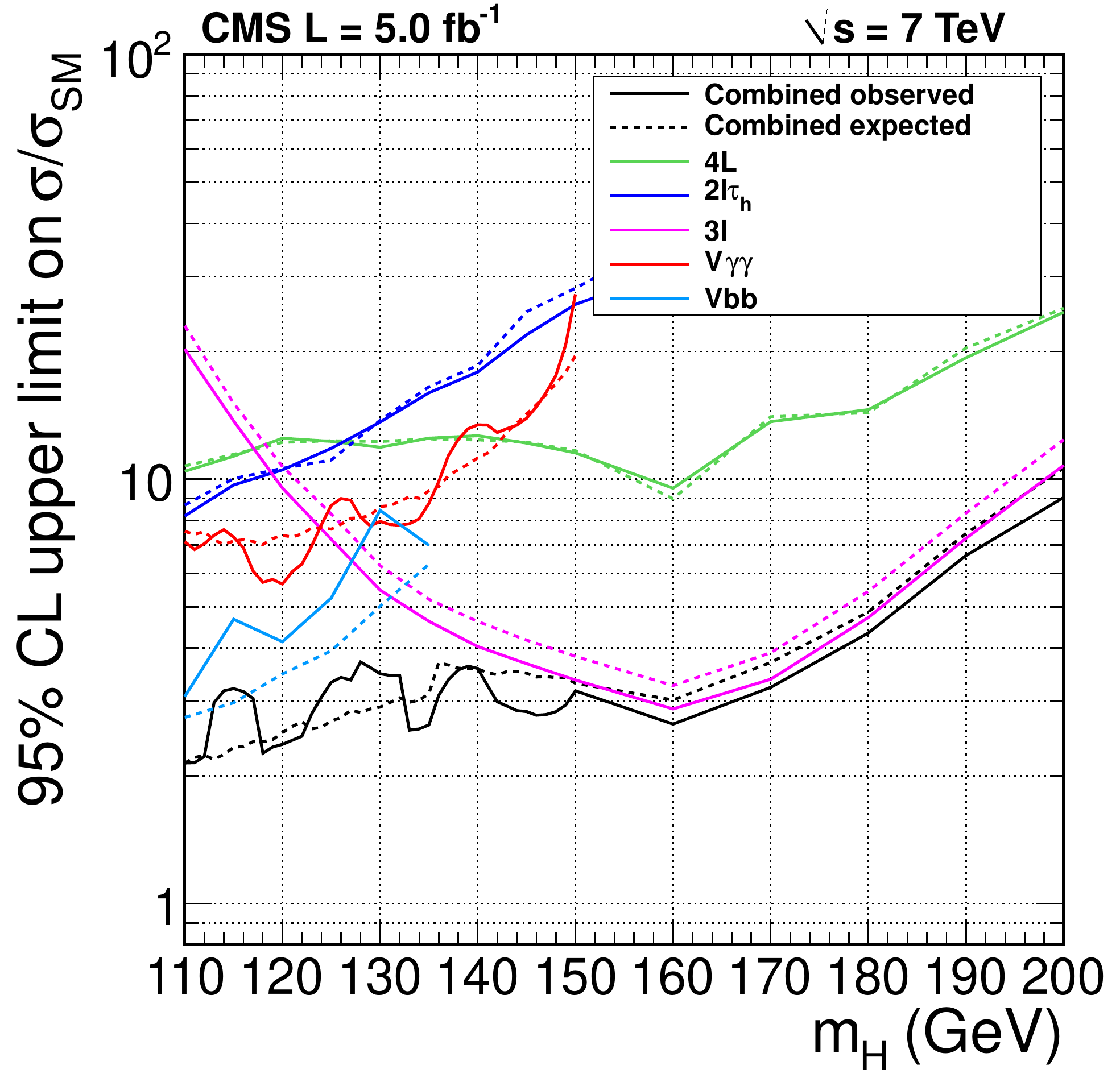}
  \caption{At left, the observed and expected limits, at 95\% CL, on SM Higgs boson production combining the AP searches presented in this paper with the previously published AP $\Hi \to\gamma\gamma$~\cite{CMS-PAS-HIG-12-009} and $\Hi \to \bbbar$~\cite{Chatrchyan2012284} searches.
  At right, the exclusive observed and expected limits (indicated by the solid and dashed lines respectively) are shown for each sub-channel.
  }
\label{fig:megalimit}
\end{center}
\end{figure}

We additionally combine the searches presented here with the CMS AP Higgs boson searches, using the same dataset, in the $\Hi\to\gamma\gamma$~\cite{CMS-PAS-HIG-12-009} and $\Hi\to\bbbar$~\cite{Chatrchyan2012284} decay modes.
The $\Hi\to\gamma\gamma$ and $\Hi\to\bbbar$ searches are included in the limit combination for Higgs boson masses below 150\GeV and 135\GeV, respectively.
The treatment of systematic uncertainties in these channels is similar to that described in Section~\ref{sec:systematics}.
The potential contributions of the VBF and GF SM Higgs boson production mechanisms to these analyses are negligible.
The associated $\ttbar\Hi$ production mechanism contributes approximately 5\% and 14\% of the expected signal yield in the $4L$ and AP $\Hi \to \gamma\gamma$ channels, respectively.
The contributions from $\ttbar\Hi$ to the other channels are negligible.
The limits for each sub-channel and for the combination of all CMS AP searches are shown in Fig.~\ref{fig:megalimit}.
The full combination excludes, at 95\% CL, the associated production of SM Higgs bosons at 2.1--3.7 times the SM prediction for Higgs boson masses below $170\GeV$.
The observed and expected limits for a Higgs boson mass of 125\GeV are enumerated for the full combination and for each exclusive sub-channel in Table~\ref{tab:limitsForTHEMass}.
\begin{table}
  \begin{center}
    \caption{Exclusive observed and expected limits for each sub-channel and for the total combination, at 95\% CL, on SM Higgs boson production for a Higgs boson mass of 125\GeV.
    \label{tab:limitsForTHEMass}
    }
    \begin{tabular} {|c|c|c|c|c|c|c|}
      \hline
      Channel & -2$\sigma$ & -1$\sigma$ & Expected & +1$\sigma$ & +2$\sigma$ & Observed\\
      \hline
      $3\ell$              &  4.93 &  5.99 &  8.28 & 11.78 & 16.47 &  7.21\\
      $2\ell\tau_h$        &  7.95 &  8.69 & 11.09 & 15.95 & 24.74 & 11.81\\
      $4L$                 &  6.79 &  8.90 & 12.31 & 17.49 & 24.62 & 12.27\\
      $\gamma\gamma$       &  5.82 &  6.51 &  7.62 & 10.86 & 15.80 &  8.67\\
      $\bbbar$             &  2.31 &  2.92 &  3.94 &  5.73 &  8.34 &  5.25\\
      \hline
      Combined             &  1.65 &  1.89 &  2.69 &  3.79 &  5.43 &  3.32\\
      \hline
    \end{tabular}
  \end{center}
\end{table}

\section{Summary}
A search for the standard model Higgs boson, produced in association with a $\W$ or $\Z$ boson, has been described.
The search is conducted using final states with three or four isolated leptons in the entire 2011 CMS dataset.
The analysis is sensitive to associated production where the Higgs boson decays into either a $\Pgt$ pair or $\W$-boson pair.
A total of 29 events are observed, and are compatible with the background prediction.
Upper limits of about 2.6--9 times greater than the predicted value are set at 95\% CL for the product of the SM Higgs boson associated production cross section and decay branching fraction in the mass range  $110 < m_{\mathrm{H}} < 200\GeV$.
The searches presented in this paper are combined with two other CMS associated production Higgs boson searches using the $\Hi \to \gamma \gamma$ and $\Hi \to \bbbar$ decay modes.
While the inclusive combination excludes, at 95\% CL, the associated production of SM Higgs bosons at 3.3~times the SM prediction for a Higgs boson with a mass of 125\GeV, all of the exclusive limits in each decay mode, and the inclusive combined limit, are consistent with the predictions of the SM.
\section*{Acknowledgements}
We congratulate our colleagues in the CERN accelerator departments for the excellent performance of the LHC machine. We thank the technical and administrative staff at CERN and other CMS institutes, and acknowledge support from BMWF and FWF (Austria); FNRS and FWO (Belgium); CNPq, CAPES, FAPERJ, and FAPESP (Brazil); MES (Bulgaria); CERN; CAS, MoST, and NSFC (China); COLCIENCIAS (Colombia); MSES (Croatia); RPF (Cyprus); MoER, SF0690030s09 and ERDF (Estonia); Academy of Finland, MEC, and HIP (Finland); CEA and CNRS/IN2P3 (France); BMBF, DFG, and HGF (Germany); GSRT (Greece); OTKA and NKTH (Hungary); DAE and DST (India); IPM (Iran); SFI (Ireland); INFN (Italy); NRF and WCU (Korea); LAS (Lithuania); CINVESTAV, CONACYT, SEP, and UASLP-FAI (Mexico); MSI (New Zealand); PAEC (Pakistan); MSHE and NSC (Poland); FCT (Portugal); JINR (Armenia, Belarus, Georgia, Ukraine, Uzbekistan); MON, RosAtom, RAS and RFBR (Russia); MSTD (Serbia); SEIDI and CPAN (Spain); Swiss Funding Agencies (Switzerland); NSC (Taipei); ThEP, IPST and NECTEC (Thailand); TUBITAK and TAEK (Turkey); NASU (Ukraine); STFC (United Kingdom); DOE and NSF (USA).

Individuals have received support from the Marie-Curie programme and the European Research Council (European Union); the Leventis Foundation; the A. P. Sloan Foundation; the Alexander von Humboldt Foundation; the Austrian Science Fund (FWF); the Belgian Federal Science Policy Office; the Fonds pour la Formation \`a la Recherche dans l'Industrie et dans l'Agriculture (FRIA-Belgium); the Agentschap voor Innovatie door Wetenschap en Technologie (IWT-Belgium); the Ministry of Education, Youth and Sports (MEYS) of Czech Republic; the Council of Science and Industrial Research, India; the Compagnia di San Paolo (Torino); and the HOMING PLUS programme of Foundation for Polish Science, cofinanced from European Union, Regional Development Fund.

\bibliography{auto_generated}   

\cleardoublepage \appendix\section{The CMS Collaboration \label{app:collab}}\begin{sloppypar}\hyphenpenalty=5000\widowpenalty=500\clubpenalty=5000\textbf{Yerevan Physics Institute,  Yerevan,  Armenia}\\*[0pt]
S.~Chatrchyan, V.~Khachatryan, A.M.~Sirunyan, A.~Tumasyan
\vskip\cmsinstskip
\textbf{Institut f\"{u}r Hochenergiephysik der OeAW,  Wien,  Austria}\\*[0pt]
W.~Adam, E.~Aguilo, T.~Bergauer, M.~Dragicevic, J.~Er\"{o}, C.~Fabjan\cmsAuthorMark{1}, M.~Friedl, R.~Fr\"{u}hwirth\cmsAuthorMark{1}, V.M.~Ghete, J.~Hammer, N.~H\"{o}rmann, J.~Hrubec, M.~Jeitler\cmsAuthorMark{1}, W.~Kiesenhofer, V.~Kn\"{u}nz, M.~Krammer\cmsAuthorMark{1}, D.~Liko, I.~Mikulec, M.~Pernicka$^{\textrm{\dag}}$, B.~Rahbaran, C.~Rohringer, H.~Rohringer, R.~Sch\"{o}fbeck, J.~Strauss, A.~Taurok, W.~Waltenberger, G.~Walzel, E.~Widl, C.-E.~Wulz\cmsAuthorMark{1}
% ...
\vskip\cmsinstskip
\dag:~Deceased\\
1:~~Also at Vienna University of Technology, Vienna, Austria\\
% ...

\end{sloppypar}
\end{document}